\documentstyle[epsf,prl,aps,draft]{revtex}

\def\mxth{\mathsurround=0pt }
\def\xversim#1#2{\lower2.pt\vbox{\baselineskip0pt \lineskip-.2pt               
    \ialign{$\mxth#1\hfil##\hfil$\crcr#2\crcr\sim\crcr}}}

\begin{document}
\draft
\title{Determining liquid structure from the tail of the direct
correlation function}
\author{Kirill Katsov and John D. Weeks}
\address{IPST and Department of Chemistry, University of Maryland\\
College Park, Maryland 20742}
\date{\today}
\maketitle

\begin{abstract}
In important early work, Stell showed that one can determine the pair
correlation function $h(r)$ of the hard sphere fluid for all distances $r$
by specifying only the ``tail'' of the direct correlation function $c(r)$ at
separations greater than the hard core diameter. We extend this idea in a
very natural way to potentials with a soft repulsive core of finite extent
and a weaker and longer ranged tail. We introduce a new continuous
function $T(r)$ which reduces exactly to the tail of $c(r)$ outside
the (soft) core 
region and show that both $h(r)$ and $c(r)$ depend only on the ``out
projection'' of $T(r)$: i.e., the product of the Boltzmann factor of the
repulsive core potential times $T(r)$. Standard integral equation closures
can thus be reinterpreted and assessed in terms of their predictions for the
tail of $c(r)$ and simple approximations for its form suggest new closures.
A new and very efficient variational method is proposed for solving the
Ornstein-Zernike equation given an approximation for the tail of $c$.
Initial applications of these ideas to the Lennard-Jones and the hard core
Yukawa fluid are discussed.
\end{abstract}


\section{Introduction}
\label{intro}One of the many areas of current research where George Stell
has made fundamental contributions is the derivation of integral equations
to determine the pair correlation function of a uniform fluid. A number of
different integral equations have been proposed \cite{hansenmac}, often
based on the graphical and functional methods pioneered by Stell \cite
{stell64}. However, despite much effort and some impressive successes, there
has been a mixed record arising from their use in different applications.
For example, while the Percus-Yevick (PY) equation \cite{PY,percus} for a
fluid of hard spheres is quite accurate, it proved much less successful in
describing the structure of systems with longer ranged interactions such as
the Lennard-Jones (LJ) fluid \cite{levesque}. In most cases, we do not have
a deep understanding of the reasons for a particular equation's success or
failure. Part of the problem is that standard ``closures'' of the integral
equations usually introduce uncontrolled approximations made mostly for
mathematical convenience. Thus it is difficult to assess the physical
consequences of the errors introduced and the kinds of interactions for
which a particular equation is likely to be accurate.

However, as pointed out by Stell in one of his earliest papers
\cite{stell63}, there is a very simple and physically suggestive way
to interpret one of 
the most basic and successful of the integral equations, the PY equation for
hard spheres. Stell noted that one can completely determine the pair
correlation function $h(r)$ of the hard sphere fluid for all distances $r$
by specifying only the {\em tail }or {\em out part} of the{\em \ }direct
correlation function $c(r)$ (i.e., its value at separations $r>d,$ with $d$
the hard core diameter of the hard spheres). Here $h$ and $c$ are related by
the usual Ornstein-Zernike (OZ)\ equation \cite{hansenmac}. See
Sec.~(\ref{dcf}) below for precise definitions and further
discussion. If, following
OZ, one further assumes that the direct correlation function has the range
of the potential, then its out part vanishes for hard spheres. Then
the {\em core} or {\em in part} of $c(r)$ for $r<d$ can be determined
directly from 
the OZ equation and the exact condition imposed by the hard core potential
that $h(r)=-1$ for $r<d$. Stell showed that the resulting $h(r)$ computed
from the OZ equation is identical to the PY solution for hard spheres.
However this simple picture directly applies only to the PY equation for
hard spheres.

Stell and other workers \cite{hoystell77} generalized this idea to apply to
potentials with a hard core and a longer ranged tail by making simple
assumptions about the functional form of the out part of $c(r)$ and solving
the OZ equation subject to the ``core condition'' $h(r)=-1$ inside the core.
The resulting mean spherical approximation (MSA) and generalized MSA (GMSA)
equations have proved useful in a variety of applications. Madden and Rice 
\cite{smsa} showed how these ideas could be applied to systems with softer
repulsive cores with their soft MSA (SMSA) equation, though the relationship
between the original hard core condition and the treatment of soft cores,
both in the initial work and in later derivations \cite{hansenmac}, seems
(to us at least!) somewhat unclear. Most recent work on integral equation
closures has focused attention on another function, the bridge function (see
Sec. \ref{bridgefunction} below), which is not simply related to the tail of 
$c,$ and connections to the earlier work and the insights gained therein
have often not been exploited.

In this paper we show how George Stell's original ideas \cite{stell63} can
be extended in a very natural way to describe more realistic systems with
finite ranged soft core interactions and/or weaker and longer ranged
(usually attractive) interactions. While some of our conclusions have been
noted before, the general perspective and the formalism we develop is new.
It gives a unified and physically suggestive way of interpreting and
assessing many earlier approaches and ideas and suggests new and simpler
approximations. The main idea is to introduce a new continuous
function 
$T(r)$ which reduces exactly to the tail of $c(r)$ outside the (soft) core
region. We show that both $h(r)$ and $c(r)$ depend only on the ``out
projection'' of $T(r)$: i.e., the product of the Boltzmann factor of the
repulsive core potential times $T(r).$ Essentially then, we have only to
prescribe $T$ outside the core, i.e., fix the tail of $c$, to determine $h$
and $c$ everywhere. This conclusion is rigorously true for hard cores, as
noted in the original work of Stell and others \cite{stell63,hoystell77}.

We thus make direct contact with a wide class of integral equations related
to the PY equation for hard spheres and the MSA and find in a new and more
straightforward way equations related to the SMSA of Madden and Rice \cite
{smsa}. Our general approach suggests how to improve the behavior of the
SMSA equation at low densities and gives new insights into reasons for the
success of some of the most accurate integral equations, including the
reference hypernetted chain (RHNC) equation suggested by Lado \cite{lado73}
and the method of Zerah and Hansen \cite{hmsa}. Equally important, many of
the inherent limitations of all these methods are clarified.

\section{System}

We consider here the simple case of a one component uniform fluid
interacting through a spherically-symmetric intermolecular pair
potential 
$w(r)=u_{0}(r)+u_{1}(r)$, where $u_{0}$ is a harshly repulsive core potential
with {\em finite range} $\bar{\sigma}$ (so $u_{0}(r)=0$ for
$r>\bar{\sigma}$) and $u_{1}$ is a longer-ranged and more slowly
varying (usually 
attractive) potential. We will refer to a system with potential $u_{0}$
alone as the {\em reference system} and the potential $u_{1}$ as the 
{\em perturbation potential}. Though many of these ideas can be
directly applied 
to fluids with long-ranged (e.g., Coulomb) forces, several new issues arise
there that merit a more detailed discussion, and we will restrict our work
here to the case where $u_{1}(r)$ goes to zero at large $r$ faster
than $r^{-3}$. We also assume in most of the following that $u_{1}$ is
continuous, 
with at least one continuous derivative at $r=\bar{\sigma}.$ Examples of a
pair potential divided in this way are the separations proposed by Ree et
al. \cite{ree} and by Weeks et al. \cite{wca} for the LJ potential.

The local density at a distance $r$ away from a particle fixed at the origin
in a fluid with average (number) density $\rho $ is given by $\rho g(r)$,
where $g(r)$ is the radial distribution function. In the following, we will
use the notation $g(r;[w])$ to indicate the functional dependence of $g(r)$
on the pair potential $w$; the subscripts $0$ will denote the reference
system and $d$ a hard sphere system with diameter $d$. Note that $g(r)$
becomes very small in the core region $r<\bar{\sigma}$ because of the
repulsive core potential $u_{0}$. In the special case where $u_{0}$ is
replaced by a hard sphere interaction $u_{d}(r)$, then $g(r;[u_{d}+u_{1}])=0$
for all $r<d.$ Our goal is to determine quantitatively the pair correlation
function $h(r)\equiv g(r)-1$ for the uniform fluid. Important thermodynamic
and structural information are contained in $h(r)$ and its calculation has
been a major focus of research in the theory of liquids \cite{hansenmac}.

\section{Direct Correlation Function}

\label{dcf}To that end most modern approaches introduce several other
related functions. Probably the most fundamental of these is the {\em direct
correlation function} $c(r)$, defined in terms of $h(r)$ by the
Ornstein-Zernike (OZ)\ equation 
\begin{equation}
h(r_{1})=c(r_{1})+\rho \int d{\bf r}_{2}c(r_{2})h(|{\bf r}_{1}-{\bf r}
_{2}|)\,.  \label{OZ}
\end{equation}
By iterating this equation $h$ can be represented as a sum of chains of
``direct'' correlations $c$. For typical short ranged potentials, this
suggests that $c$ could be both shorter ranged than $h$ and simpler in
structure \cite{hansenmac}. Indeed, Ornstein and Zernike \cite{OZ} assumed
that $c$ had the range of the intermolecular potential in developing their
theory of correlations near the critical point. While scaling theory shows
that $c$ must in fact decay as a power law $r^{-\eta }$ at the critical
point, Stell and co-workers \cite{scoza1} have shown that very accurate
results can be obtained for thermodynamic properties of the lattice gas
surprisingly close to the critical point by assuming $c$ is strictly the
range of the potential and choosing its form to yield self-consistent
thermodynamic predictions. Moreover, for the long ranged Coulomb potential,
assuming that $c$ is proportional to the potential physically incorporates
the effects of screening and yields a nonlinear version of Debye-H\"{u}ckel
theory \cite{hansenmac}.

We refer to the idea that $c$ has (to a good approximation) the range of the
potential as the {\em range assumption.} A very direct but primitive
strategy for calculating $h$ is to guess the form of the presumably simpler
function $c,$ perhaps guided by the range assumption, and then determine $h$
from the OZ equation. However, Stell's interpretation of the PY equation for
hard spheres \cite{stell63} suggests a simpler possibility: perhaps we have
to prescribe only the {\em tail} of $c$ outside the range of the harshly
repulsive core potential $u_{0}$ to determine $h.$ We now develop a general
formalism incorporating this idea for a system with potential 
$w(r)=u_{0}(r)+u_{1}(r)$.

\section{Core and Tail Projections Using Continuous Functions}

\label{coreandtail}To help us focus on the core and tail parts of functions,
we note that the Boltzmann ($e_{0}$) and Mayer ($f_{0}$) functions for the
harshly repulsive core potential $u_{0}(r)$ act very nearly as 
{\em projection operators} onto {\em tail} or {\em out} ($r>\bar{\sigma}$) and 
{\em core} or {\em in} ($r<\bar{\sigma}$) subspaces respectively, since

\begin{equation}
e_{0}(r)\equiv e^{-\beta u_{0}(r)}\; 
\begin{array}{cc}
\approx 0, & r<\bar{\sigma}, \\ 
=1, & r>\bar{\sigma},
\end{array}
\hspace{0.2in}-f_{0}(r)\equiv 1-e^{-\beta u_{0}(r)}\; 
\begin{array}{cc}
\approx \;1, & r<\bar{\sigma}, \\ 
=0, & r>\bar{\sigma}.
\end{array}
\;.
\end{equation}

These functions exactly satisfy one property of orthogonal projectors for
all $r$ : 
\begin{equation}
-f_{0}(r)+e_{0}(r)=1,  \label{fpluse}
\end{equation}
and in the tail region $r>\bar{\sigma}$ exactly satisfy the second
requirement: 
\begin{equation}
-f_{0}(r)\cdot e_{0}(r)=0.  \label{ftimese}
\end{equation}
Moreover for small $r<$ $\bar{\sigma}$ well inside the core, the repulsive
potential $u_{0}$ is very large and $e_{0}$ essentially vanishes. Thus Eq.~(%
\ref{ftimese}) also holds in this region to a very good approximation.

However, for soft cores there is a {\em transition region} for $r$ near $%
\bar{\sigma}$ where the r.h.s. of Eq.~(\ref{ftimese}) differs significantly
from zero. Thus strictly speaking the functions $-f_{0}$ and $e_{0}$ are not
true projection operators over all space. Rather they divide space into two
parts: a tail or out part, and a core or in part. The latter is comprised of
a transition region for $r$ near $\bar{\sigma}$ and an effective hard core
region at smaller $r$. The theory for soft cores we develop works best when
the spatial extent of the transition region is much smaller than 
$\bar{\sigma}$, as is the case for harshly repulsive interactions. In
the special case 
where there is a hard core potential $u_{d}$, the width of the transition
region vanishes, Eq.~(\ref{ftimese}) holds exactly for all $r,$ and the
corresponding functions $-f_{d}$ and $e_{d}$ are true projection operators.
Our theory for soft cores will go over smoothly to that for hard cores in
the limit of increasing steepness of the soft core potential.

We now rewrite our correlation functions in projected form. Though our
primary focus has been on the pair of functions $h$ and $c,$ both have
discontinuities at $r=d$ when there is a hard core potential $u_{d}.$ It is
convenient to introduce two new functions that remain continuous even in
this limit and from which we can determine both $h$ and $c$. One such
function we will use is well known and was originally used by Stell \cite
{stell63}: 
\begin{equation}
t(r)\equiv h(r)-c(r).  \label{t definition}
\end{equation}
$t$ is sometimes referred to as the ``indirect correlation function'' 
\cite{lee95}; its continuity even when the potential has a hard core region is
clear since it equals the convolution integral in the OZ equation
(\ref{OZ}). From this it follows that the first $D$ derivatives of $t$
in a $D$-dimensional system are also continuous at $r=d$ even for a hard core
system. For harshly repulsive core potentials it is easy to relate $c$
for 
$r<\bar{\sigma}$ to the core part of $t$: to a very good approximation in the
effective hard core region we have 
\begin{equation}
c(r)\approx f_{0}(r)[1+t(r)],\;\;r<\bar{\sigma}.  \label{cinapprox}
\end{equation}
This equation is exact for a hard core potential where $f_{d}$ and $h_{d}=-1$
for all $r<d.$

To determine $c$ outside the core, we now introduce a second continuous
function, which we refer to as the {\em tail function} $T(r)$, whose out
projection $e_{0}(r)T(r)$ reduces {\em exactly} to the tail of $c$ in the
out region. In the core space we require that $e_{0}(r)T(r)$ correct the
small errors in Eq.~(\ref{cinapprox}) occurring in the transition region
for soft cores. Thus we require for all $r$ that $T(r)$ satisfy: 
\begin{equation}
c(r)=f_{0}(r)[1+t(r)]+e_{0}(r)T(r).  \label{cSinout}
\end{equation}
Moreover, since $g=c+1+t,$ we have, using Eqs.~(\ref{fpluse}) and (\ref
{cSinout}) 
\begin{equation}
g(r)=e_{0}(r)[1+t(r)]+e_{0}(r)T(r).  \label{gSD}
\end{equation}
We have thus rewritten $c$ and $g$ (or $h)$ in projected form using the new
functions $t$ and $T.$ While special cases of these equations have been
suggested before \cite{stell63}, the general utility of such a $T$ function
does not seem to have been realized. The most important properties of the
tail function $T$ are clear from Eqs.(\ref{cSinout}) and (\ref{gSD}): i) it
reduces {\em exactly} to the tail of $c$ in the out region; ii) {\em
both} 
$h$ and $c$ depend on $T$ only through the combination $e_{0}T$; iii) $T$ is
continuous and differentiable.

To see that the latter holds, let us define the {\em cavity distribution
function} $y(r)$ in the usual way \cite{hansenmac} : $y(r)\equiv e^{+\beta
w(r)}g(r)$. Simple analysis like that mentioned above for $t(r)$ (see, e.g.,
Ref. \cite{percus}) shows that $y(r)$ is a well-defined continuous function
of $r$ with several continuous derivatives even when $w$ itself has a hard
core region or other discontinuities. Using Eq.~(\ref{gSD}) we immediately
get that 
\begin{equation}
y(r)=[1+t(r)+T(r)]/e_{1}(r).  \label{yofr}
\end{equation}
Here $e_{1}(r)\equiv e^{-\beta u_{1}(r)}$. Since $y(r)$ and $t(r)$ are
continuous and differentiable and the perturbation tail function $e_{1}(r)$
can be constructed to be continuous and differentiable even across a hard
core region, it follows that $T(r)$ is continuous and differentiable 
\footnote{More generally, we can exploit the fact that $y$ and $t$
have at least 2 continuous derivatives for $D=3$, to relate the behavior of
low order derivatives of $T$ to those of $e_{1}.$ This could be used to give
a more accurate extrapolation of $T$ into the transition
region.}. 
When the potential has a hard core, Eq.~(\ref{yofr}) can 
alternatively be used to define $T(r)$ for all $r$ in terms of the more
familiar functions $y,$ $t,$ and $e_{1}.$

\section{Basic Result}

\label{dependence}Now we can refine the primitive strategy of guessing $c$
and using the OZ equation to calculate $h,$ by reexpressing everything in
terms of $t$ and $T.$ See the Appendix for numerical details. In principle,
if we prescribe $T(r)$ for all $r$ then $t(r)$ can be completely determined
from the modified OZ equation. However, we see from Eqs.~(\ref{cSinout}) and
(\ref{gSD}) that since both{\em \ }$g$ and $c$ (and hence also $t$) depend
only on $e_{0}T,$ the results are very insensitive to any errors we make in
prescribing $T$ in the core space $r<\bar{\sigma}$. This is obvious in the
effective hard core region where $e_{0}$ essentially vanishes. In the
narrow\ transition region, since $T$ is continuous and differentiable, its
values there can be accurately determined by {\em extrapolation} from those
for $r\gtrsim \bar{\sigma}.$ In effect then we only have to prescribe the 
{\em out part} of $T,$ i.e., the {\em tail} of $c$, to determine both $h$
and $c$ everywhere. This generalizes Stell's argument \cite{stell63} for the
hard core PY equation. In the Appendix we introduce a new and very efficient
variational method that allows us to determine numerically both $h$ and $c$
from the OZ equation given some approximation for the out part of $T(r).$
This will allow us to find accurate solutions to many standard integral
equations in a very simple way.

Note from Eq.~(\ref{yofr}) that the tail of $c$ is {\em not} sufficient to
determine $y(r).$ Its values for small $r$ in the effective hard core region
depend directly on $T(r)$ there and we cannot expect that extrapolation from
the out part of $T$ alone will give accurate results for $T(r)$ well inside
the core. From this perspective, the calculation of $y(r)$ (and other
closely related functions such as the bridge function $B(r)\equiv \ln
y(r)-t(r)$ discussed below in Sec. \ref{bridgefunction})\ is a much more
difficult problem, requiring the accurate determination of {\em both} the
out and core parts of $T(r).$ Fortunately the latter problem does not have
to be solved to find accurate results for $h$ and $c.$ This point was
emphasized by Stell for the hard sphere system \cite{stell63}, and we see it
holds true much more generally.

\section{Relation to Previous Work}

\label{previouswork}Stell's original work \cite{stell63} was designed to
provide information about the PY equation for a system with the general pair
potential $w(r).$ To that end, he introduced a set of equations very similar
in form to Eqs.~(\ref{cSinout}), (\ref{gSD}), and (\ref{yofr}), but with the
crucial difference that the Boltzmann and Mayer functions $e$ and $f$ for
the {\em full} potential $w$ appear, where

\begin{equation}
e(r)\equiv e^{-\beta w(r)}=e_{0}(r)e_{1}(r)\;;\hspace{0.2in}f(r)\equiv
e^{-\beta w(r)}-\;1=f_{0}(r)+e_{0}(r)f_{1}(r)\;.  \label{fulleandf}
\end{equation}
Here $f_{1}(r)\equiv e^{-\beta u_{1}(r)}-\;1$. Note that $f$ has the range
of the full potential and $-f$ and $e$ no longer approximate projection
operators onto core and tail regions. Stell's equations can be written as 
\begin{equation}
c(r)=f(r)[1+t(r)]+e(r)d(r),  \label{cStell}
\end{equation}
\begin{equation}
g(r)=e(r)[1+t(r)]+e(r)d(r),  \label{gStell}
\end{equation}
\begin{equation}
y(r)=1+t(r)+d(r).  \label{yStell}
\end{equation}
Eq.~(\ref{yStell}) can be taken as the definition of the function $d(r)$ (we
use Stell's notation; this should not be confused with the hard sphere
diameter). Despite the superficial similarity of these equations to our
Eqs.~(\ref{cSinout}), (\ref{gSD}), and (\ref{yofr}), $d(r)$ in general has
very different properties than our analogous function $T(r).$ In particular, 
$d(r)$ does not reduce to the tail of $c$ in the out region and is likely to
have a more complicated oscillatory structure. The main utility of Eqs.~(\ref
{cStell}), (\ref{gStell}), and (\ref{yStell}) is in analyzing the PY
equation: Stell was able to show that the usual formulation of the PY
equation for a general potential results from the approximation $d(r)=0$.
Unfortunately there is little reason to believe this approximation is
generally accurate.

However, in the special case of hard core interactions where $w(r)=u_{d}(r)$
, Eqs.~(\ref{cStell}), (\ref{gStell}), and (\ref{yStell}) reduce to
our 
Eqs.(\ref{cSinout}), (\ref{gSD}), and (\ref{yofr}), and
$d_{d}(r)=T_{d}(r).$ The 
approximation $d_{d}(r)=0$ in the out region for hard spheres then can be
motivated by an application of the range assumption for the tail of $c$.
This assumption alone is enough to determine the accurate PY solution
for 
$h_{d}(r)$. The range ansatz $d_{d}(r)=0$ for $r>d$ is {\em exact} in one
dimension ($D=1$) and hence yields the exact $h_{d}(r)$. In $D=3$, the first
errors in $h_{d}^{PY}(r)$ show up at $O(\rho ^{2})$ in a density expansion.
Overall $h_{d}^{PY}(r)$ remains remarkably close to the results of computer
simulations even at higher densities, with small errors most noticeable near
contact and at the first minimum for densities near the fluid-solid
transition \cite{hansenmac}. As noted by Stell \cite{stell63}, all that is
required to calculate $h_{d}(r)$ in general is an expression for $d_{d}(r)$
in the out region. Essentially exact results for $h_{d}(r)$ can be obtained
from the {\em generalized MSA} (GMSA) of Waisman and Lebowitz \cite{wais},
which assumes the existence of a small short-ranged (Yukawa-like) tail
in 
$c_{d}(r)$ for $r>d$. Parameters in the tail are chosen so that $h_{d}$ gives
results for the pressure and compressibility that fit simulation data. The
basic picture suggested by the range assumption that the tail of $c_{d}$ has
a simple structure and is small and much shorter ranged than $h_{d}$ seems
to be well established.

Stell \cite{stell63} also noted that the extrapolation of the PY
approximation $d_{d}(r)=0$ deep into the core space is a separate and much
less accurate approximation. For example, the resulting PY expression
for 
$y_{d}(r)$ given by Eq.~(\ref{yStell}) with $d_{d}(r)=0$ for all $r<d$ can
have large errors at small $r$ for $D=1$ even though the PY result for
$h_{d}(r)$ is exact. (While $d_{d}(r)$ is continuous and
differentiable at 
$r=d$ higher derivatives are discontinuous, leading to a large positive value
at small $r$ for the exact $d_{d}(r)$ at high density.) This strongly
suggests that the calculation of $h_{d}$ and $y_{d}$ should be logically
separated \cite{contrary}. Of course, $y_{d}$ is an interesting function and
additional properties like the chemical potential can be obtained from it 
\cite{hansenmac}. However, a focus on $h_{d}$ and $c_{d}$ alone permits a
very simple theory, and one can use results for the pressure and
compressibility from $g_{d}(r)$ and $c_{d}(r)$ and thermodynamic relations
to calculate other thermodynamic properties. In particular, in this approach
the chemical potential should be calculated by integrating the pressure, and
not from the very inaccurate value for $y_{d}(0)$ given by
extrapolating 
$d_{d}(r)=0$ deep into the core space. By introducing the tail function $T(r)$
and the system of equations (\ref{cSinout}), (\ref{gSD}), and (\ref{yofr}),
we have been able to extend these important ideas of Stell for hard sphere
systems \cite{stell63} to systems with more general interactions.

\section{General Properties of the Tail Function}

\label{generalprop}We now describe some general properties of $T(r)$. Using
Eqs.~(\ref{cSinout}), (\ref{gSD}), and (\ref{yofr}), this can be rewritten
exactly as 
\begin{equation}
T(r)=c(r)-f_{0}(r)e_{1}(r)y(r),  \label{Tyandc}
\end{equation}
explicitly showing that $T$ reduces to the tail of the direct correlation
function in the out region, but has a different form in the core region. To
focus on the changes induced by the perturbation potential $u_{1}$, it is
useful to define the {\em excess quantities}: 
\begin{equation}
\Delta T(r)\equiv T(r)-T_{0}(r),  \label{T0T1}
\end{equation}
where $T_{0}$ is the exact $T$ function for the reference system, with
similar definitions for other excess functions such as $\Delta h$ and 
$\Delta c.$ According to the range ansatz $T_{0}$ is zero in the out region,
and we expect that the exact $T_{0}$ will in general be small and vanish
rapidly at larger $r$ outside the core. Thus in the out region $T(r)\approx$
$\Delta T(r),$ and is mainly determined by the potential tail $u_{1}(r)$.

Based on an analysis by Stell \cite{stell77}, it is generally believed that
away from the critical point the asymptotic form of $c(r)$ at large $r$ is

\begin{equation}
c(r)\sim -\beta u_{1}(r).  \label{climit}
\end{equation}
For system with a weak and slowly varying potential tail $u_{1}$ that goes
smoothly to zero at large $r$ this is consistent with the idea that the OZ
equation should reduce to linear response theory far from the core region.
Here $\beta \ $is the inverse of Boltzmann's constant times the temperature.
Thus we expect $\Delta T(r)\sim -\beta u_{1}(r)$ far from the core.

At very low density $\rho $ graphical expansion methods show that the exact
form of $c(r)$ for interaction potentials going to zero faster than $r^{-3}$
can be written as: 
\begin{equation}
c(r)=f(r)[1+\rho \Lambda (r)]+O(\rho ^{2}),  \label{clow}
\end{equation}
where 
\begin{equation}
\Lambda (r_{12})=\int d{\bf r}_{3}f(r_{13})f(r_{32}).  \label{lambda}
\end{equation}
Note that the range assumption for $c$ is rigorously true at low density.
Similarly it is easy to show that 
\begin{equation}
t(r)=\rho \Lambda (r)+O(\rho ^{2}),  \label{tlow}
\end{equation}
\begin{equation}
y(r)=1+\rho \Lambda (r)+O(\rho ^{2}),  \label{ylow}
\end{equation}
and 
\begin{equation}
T(r)=f_{1}(r)[1+\rho \Lambda (r)]+O(\rho ^{2}).  \label{Tlow}
\end{equation}
It follows from Eq.~(\ref{Tlow}) that $T_{0}(r)=0+O(\rho ^{2})$.

\section{Closures and the Tail Function}

\label{closures}Most integral equation theories for $h(r)$ are based on the
idea of a {\em closure }\cite{hansenmac}: a second relation between
$h$ and 
$c$ which, when combined with the OZ equation, allows one to solve for the
values of $h$ and $c$. However most closures are expressed in terms of more
complicated functions like $y(r)$ or $B(r)$ and their form is usually
determined by mathematical considerations. See, e.g., Sec.~(\ref{unique})
below. The above results show that to calculate $h(r)$ we can focus on the
simpler projected function $e_{0}(r)T(r),$ determined essentially only by
the tail of $c(r).$ An exact choice will yield an exact $h$ and approximate
choices can be motivated by the range ansatz and the general supposition
that the tail of $c$ has a simple structure. As discussed in the Appendix,
we can also exploit the relatively simple nature of the out part of $T(r)$
in the numerical solution of the resulting integral equations. Other
standard closures can be reinterpreted and sometimes simplified by looking
at their predictions for the tail of $c.$

\subsection{Soft Mean Spherical Approximation}

\label{SMSA}Probably the simplest such prediction directly yields the SMSA
integral equation \cite{smsa}. The SMSA assumes that the limiting linear
response value for the tail of $c$ given in Eq.~(\ref{climit}) holds for all 
$r$ in the out region. Thus we set 
\begin{equation}
e_{0}(r)T^{SMSA}(r)=e_{0}(r)[-\beta u_{1}(r)]  \label{Tsmsa}
\end{equation}
in Eqs.~(\ref{cSinout}) and (\ref{gSD}). In the out region we have $%
T_{0}^{SMSA}=0$ and $\Delta T^{SMSA}=-\beta u_{1}(r)$. The resulting
expressions for $h$ and $c$ can easily be shown to be equivalent to the
original SMSA results, which were written in a different form. If $u_{1}=0$
then the SMSA reduces to the PY equation for the reference system$.$ The
approximation $T_{0}^{SMSA}=0$ in the out region again can be motivated by
the range assumption. When $u_{0\text{ }}$is replaced by a hard core
potential $u_{d}$ then Eqs.~(\ref{cSinout}) and (\ref{gSD}) with Eq.~(\ref
{Tsmsa}) reduce to the original hard core MSA. This derivation and
interpretation of the SMSA and its relation to the MSA seems much simpler
than that found in previous work.

One way to improve the SMSA is to improve its description of repulsive
forces. Equation (\ref{Tsmsa}) sets $T_{0}^{SMSA}=0$ in the out region. If a
more accurate expression for $T_{0}$ is known this could be used along with
the MSA approximation $\Delta T^{SMSA}=-\beta u_{1}$ in the r.h.s.~of Eq.~(%
\ref{Tsmsa}). For hard cores the GMSA \cite{wais} should give a very
accurate expression for $T_{d}(r).$ Its use in the r.h.s.~of Eq.~(\ref{Tsmsa}%
) for a system with potential $w=u_{d}+u_{1}$ would yield a theory
essentially equivalent to the optimized random phase (ORPA) theory of
Andersen and Chandler \cite{ac72}, where exact hard sphere correlation
functions are supposed to be used along with a MSA treatment of $u_{1}$.

The SMSA gives rather accurate results for the high density LJ fluid and
correctly describes the qualitative changes in $\Delta h\equiv h-h_{0}$
induced by $u_{1}.$ However, it is much less accurate at low densities. This
can be understood since Eq.~(\ref{Tsmsa}) does not reduce to the exact
result, Eq.~(\ref{Tlow}), at low densities. An improved theory would result
from approximations for $T(r)$ that interpolate between the exact low
density limit, Eq.~(\ref{Tlow}), and Eq.~(\ref{Tsmsa}) at high density. We
will describe several such theories below.

\subsection{PY and HNC Equations}

\label{PYHNC}Other integral equation closures can be reexpressed in terms of
their predictions for the out part of $T$. In many cases this can give us
insights into their strengths and weaknesses. For example, by rewriting the
standard expression $g^{HNC}=\exp (-\beta w+t)$ given by the hypernetted
chain (HNC) equation \cite{hansenmac} in the projected form of
Eq.~(\ref{gSD}), we find that the HNC closure predicts 
\begin{equation}
T^{HNC}=\exp (-\beta u_{1}+t)-(1+t).  \label{Thnc}
\end{equation}
This agrees with the exact Eq.~(\ref{Tlow}) at low density. However, when
applied to the reference system, Eq.~(\ref{Thnc}) predicts that $%
T_{0}^{HNC}=\exp (t_{0})-(1+t_{0}).$ Since $t_{0}$ is large and oscillatory
at higher density in the out region, this strongly violates the range
assumption. Indeed the HNC equation gives very poor results for a dense hard
sphere system. Experience has shown that the HNC closure does a much better
job of describing slowly varying interactions, and for systems with
long-ranged Coulomb forces it is often the theory of choice \cite{hansenmac}%
. As discussed below one of the most accurate integral equation theories,
the RHNC theory \cite{lado73}, combines a HNC treatment of the more slowly
varying potential $u_{1}$ along with an (in principle) exact treatment of
reference system correlations.

The PY closure for the reference system incorporates the range assumption
and gives a much better description of reference system correlations than
does the HNC. However, for the full system it predicts for the out part of $T$:

\begin{equation}
T^{PY}=f_{1}(1+t).  \label{py}
\end{equation}
This again agrees with Eq.~(\ref{Tlow}) at low density. However at higher
density the oscillations in $t$ and the strong nonlinear dependence on the
perturbation potential will yield a larger and more oscillatory tail for $c$
than suggested by the SMSA in Eq.~(\ref{Tsmsa}). In practice the simple
linear response form of the SMSA gives much more accurate results at high
density \cite{smsa}.

\subsection{Bridge Function}

\label{bridgefunction}Most recent integral equation closures focus attention
on another continuous and differentiable function, the {\em bridge function} 
$B(r)$, which can be defined formally as \cite{hansenmac} 
\begin{equation}
B(r)\equiv \ln y(r)-t(r).  \label{bridge}
\end{equation}
Thus $g(r)\equiv \exp [-\beta w(r)+t(r)+B(r)].$ $B(r)$ represents the sum of
a well-defined set of Mayer cluster diagrams, and the HNC equation results
from the approximation $B(r)=0$. $B$ plays a role analogous to our function $%
T$ in generating closures, and we shall see that some of its relevant
properties can be understood more easily from those of $T.$ Thus one can
represent $h$, $c$, and $y$ in terms of the pair of functions $B$ and $t$.
If $B$ is specified by some closure ansatz, then these functions can be
calculated using the OZ equation.

Alternatively, using Eqs.~(\ref{cSinout}), (\ref{gSD}), and (\ref{yofr}), we
can exactly express $B$ in terms of $t$ and $T$: 
\begin{equation}
B(r)=\ln [1+t(r)+T(r)]-[t(r)-\beta u_{1}(r)].  \label{BandT}
\end{equation}
Thus $B$ depends on $T$ itself rather than the projected function $e_{0}T$,
and in that sense is a more complicated function than $h$ or $c.$ Indeed
determining its form, particularly inside the core, has proved a very
difficult challenge both for theory and simulation, and definitive results
are still not known \cite{duhhay}. However, since the out part of $B$ in
Eq.~(\ref{BandT}) can determine the out part of $T$, we can effectively
concentrate only on the out part of $B$ if we restrict ourselves to theories
for $h$ and $c.$

In general, the out part of $B$ has a rather complicated oscillatory
structure. For example, for the reference system we have exactly in the out
region, using the definition of $t$, and the equality of the tails of $c$
and $T,$%
\begin{equation}
B_{0}(r)=\ln [1+h_{0}(r)]-h_{0}(r)+T_{0}(r),\;\;r>\bar{\sigma}.
\label{B0out}
\end{equation}
Since the exact $T_{0}$ is almost certainly small and very short ranged, as
suggested by the range ansatz and the success of the PY equation for
repulsive forces, $B_{0}$ will have longer ranged oscillations determined by
those of the pair correlation function $h_{0}$. Setting $T_{0}=0$ in Eq.~(%
\ref{B0out}) yields the PY expression for the reference system bridge
diagrams.

However, in many cases the oscillatory tail of $B(r)$ for the full system
seems to depend only weakly on the perturbation potential $u_{1}(r),$ so
that $B(r)$ $\approx $ $B_{0}(r).$ This idea has been called the {\em %
universality of the bridge function} \cite{universal},{\em \ }with $B_{0}$
often approximated by $B_{d}$, the bridge function of an appropriately
chosen hard sphere system 
\footnote{In applications to the RHNC equation, discussed
in Sec. \ref{RHNC}, the hard sphere diameter $d$ is often taken as a
parameter that can be varied to achieve more consistent thermodynamic
predictions from the full system's correlation functions. However, for the
systems we consider here with short-ranged interactions, it seems more
realistic to fit $d$ to properties of the reference system using, say, the
blip function expansion \cite{hansenmac}. For more accuracy, one can
directly approximate $B_{0}$ using various closures that accurately describe
soft repulsive systems, as suggested in Ref. \cite{ree}. Coulomb systems
with strong long-ranged repulsive and attractive forces require special
treatment, and can have correlation functions differing considerably from
those of the reference system. In such cases, choosing $d$ to represent some
effective hard core diameter for the full system may be a reasonable first
approximation.}.
The following
argument gives some insight into why this could be a reasonable
approximation for the {\em out part} of $B.$ Analogous to Eq.~(\ref{B0out}%
), we have exactly 
\begin{equation}
B(r)=\ln [1+h(r)]-h(r)+[T(r)+\beta u_{1}(r)],\;\;r>\bar{\sigma}.
\label{Bout}
\end{equation}
At high density, the structure is dominated by repulsive forces for systems
with short-ranged interactions \cite{wca} and it is a fairly good
approximation to set $h(r)\approx h_{0}(r)$ (``universality'' of the
correlation functions!) Moreover the success of the SMSA suggests that $%
T(r)\approx -\beta u_{1}(r)$ and $T_{0}(r)\approx 0$ are also reasonable
approximations in the out region. Then Eqs.~(\ref{B0out}) and (\ref{Bout})
yield $B(r)\approx B_{0}(r)$ in the out region. Note that this result is 
{\em exact} at low density since $B=0+O(\rho ^{2}).$ Thus for this class of
systems, we can arrive at the idea of approximate bridge function
universality outside the core using the more physically transparent
arguments of the SMSA. Differences in the results for the two theories
should be small at high density. It can be seen using the general expression
for $B$ in Eq.~(\ref{BandT}) that these arguments do not hold for the core
part of $B$ and we see no reason to expect any such ``universality'' at
higher densities there.

\subsection{RHNC Equation}

\label{RHNC}Alternatively, if we assume it is a good approximation to set $%
B(r)\approx B_{0}(r)$ in the out region, then for systems where $h(r)\approx
h_{0}(r)$ we have $T(r)\approx -\beta u_{1}(r)$ from Eqs.~(\ref{B0out}) and (%
\ref{Bout}), which is the SMSA closure. At low density $h(r)\approx h_{0}(r)$
is not accurate, and the true $T(r)$ must differ significantly from the SMSA
prediction. Indeed using the exact low density forms for $h$ and $h_{0}$
along with $B(r)=B_{0}(r)$ in Eqs.~(\ref{B0out}) and (\ref{Bout}) yields the
exact low density form for $T$ given in Eq.~(\ref{Tlow}). Thus a theory
incorporating $B(r)\approx B_{0}(r)$ in the out region will give exact
results for $h$ at low density and should give results at high density close
to those of the accurate SMSA.

This is what is done in the RHNC theory of Lado \cite{lado73}, and overall
this is one of the most successful integral equation methods known. The
standard RHNC closure can be written as 
\begin{equation}
g^{RHNC}(r)=\exp [-\beta w(r)+t(r)+B_{0}(r)],  \label{grhnc}
\end{equation}
thus replacing the exact bridge function $B$ by $B_{0}.$ To describe its
predictions in terms of $T,$ it is convenient to consider excess functions
like that defined in Eq.~(\ref{T0T1}). We find 
\begin{equation}
e_{0}(r)\Delta T^{RHNC}(r)=g_{0}(r)\{\exp [-\beta u_{1}(r)+\Delta
t(r)]-1\}-e_{0}(r)\Delta t(r).  \label{Trhnc}
\end{equation}
Note that we only require accurate values for $g_{0}(r)$ and not for $%
B_{0}(r)$ well inside the core to determine this fundamental quantity in our
approach 
\footnote{An even simpler approximation in the spirit of the
RHNC equation suggests itself, where $g_{0}(r)$ in Eqs.~(\ref{Trhnc}) and (%
\ref{xirhnc}) is replaced by $e_{0}(r).$ We expect this to have essentially
the same behavior at both high and low density.}.
A numerical solution can be found using the
general variational method described in the Appendix.

To examine the relation between the RHNC and the SMSA more quantitatively,
let us define 
\begin{equation}
\Delta T(r)\equiv -\beta u_{1}(r)+\xi (r).  \label{xi}
\end{equation}
For the SMSA $\xi ^{SMSA}(r)=0$ in the out region. We find that in the out
region $\xi ^{RHNC}(r)$ can be written exactly as 
\begin{equation}
\xi ^{RHNC}(r)=\Delta h(r)-\ln [\Delta h(r)/g_{0}(r)+1],\;\;r>\bar{\sigma}.
\label{xirhnc}
\end{equation}
This agrees with exact results from Eqs.~(\ref{Tlow}) at low density and
corrects the poor behavior of the SMSA there. At higher density, $\xi
^{RHNC} $ represents an additional oscillatory component in the tail of $T$
when compared to the SMSA.. However, when $\Delta h$ is small, as is
generally the case at high density for the systems we consider, then $\xi
^{RHNC}$ is small (with $\xi ^{RHNC}$ vanishing whenever $\Delta h(r)=0$).
Thus $\Delta T^{RHNC}\approx \Delta T^{SMSA}=-\beta u_{1}$ in the out region
at high density, as argued above.

\subsection{Unique Function Ansatz}

\label{unique}Several workers have tried to find more accurate expressions
for $B(r)$ by assuming it is some {\em unique local function} \cite{lee92}
of $t(r),$ as suggested by several approximate closures that gave good
results for systems with short ranged repulsive interactions \cite
{hsclosures}. LLano-Restrepo and Chapman (LC) showed for systems with an
attractive potential tail $u_{1}$ that this assumption was generally
inaccurate at small $r$ in the core region and also was inaccurate at high
density outside the core \cite{LC}$.$ They proposed that there could exist
some ``renormalized'' function $\tilde{t}(r)$ involving $u_{1}$ such that $B$
is a local function of $\tilde{t}.$ They found that the choice 
\begin{equation}
\tilde{t}(r)=t(r)-\beta u_{1}(r)  \label{ttilde}
\end{equation}
gave accurate results at high density in the out region for the LJ fluid.
This is precisely what would have been suggested by applying the SMSA
closure $T(r)=-\beta u_{1}(r)$ to the exact Eq.~(\ref{BandT}) in the out
region.

However, the SMSA approximation for $T$ is not accurate well inside the core
space and indeed the renormalized function gave poor results there. Moreover
the SMSA approximation for $T$ is inaccurate at low density where the
exact $T$ reduces to $f_{1}.$ Indeed this shows that the local
function ansatz for $B$ cannot in general be correct even outside the
core. Duh and Haymet \cite {duhhay} and Duh and Henderson
\cite{duhhen} have proposed different density dependent separations of
the total potential: 
$w(r)=\tilde{u}_{0}(r;\rho)+\tilde{u}_{1}(r;\rho)$, 
with ``reference'' ($\tilde{u}_{0}$) and ``perturbation''
($\tilde{u}_{1}$) parts chosen such that Eq.(\ref{ttilde}), now
defined with $\tilde{u}_{1},$ could give more accurate results for $B$ 
as a local function of $\tilde{t},$ even well inside the core where LC's
original suggestion most noticeably failed. It is clear from
Eq.~(\ref{BandT})
that the unique function ansatz can give exact results at low density only
if the perturbation $\tilde{u}_{1}(r;\rho )$ vanishes as $\rho \rightarrow 0$
since then $T\rightarrow T_{0}=0,$ as shown in Refs. \cite{duhhay} and \cite
{duhhen}. Assessing the nature of errors induced by the unique function
approximation in general remains a very difficult task. For our purposes
here it seems simpler and more direct to retain the original physically
motivated separation and focus instead on the out part of $T,$ whose density
dependence is such that $T$ reduces to $f_{1}$ at low density while
approximating $-\beta u_{1}$ at high density.

\section{Closures Satisfying Consistency Conditions}

A natural idea is to consider more general density-dependent expressions for 
$T$ that can vary between these limits, as suggested by the RHNC equation.
Parameters in the interpolation function can be chosen to fit simulation
data or to satisfy various thermodynamic consistency conditions (Maxwell
relations and sum rules) which the exact correlation functions must obey. We
first discuss one of the most successful integral equation approaches, the
method of Zerah and Hansen (ZH) from this perspective \cite{hmsa}, and then
introduce a new and simplified method which implements this idea in a very
direct fashion. Results seem very promising. Contact is also made with very
recent work by Stell and coworkers \cite{scoza2}.

\subsection{HMSA Equation.}

\label{ZH}ZH introduced a generalized ``HMSA'' closure that interpolates
nonlinearly between the SMSA closure at small $r$ and the HNC closure at
large $r$, with a parameter in the interpolation function chosen to give
consistent results for the pressure computed from the virial and
compressibility formulas \cite{hmsa}. The choice of the HNC theory at large
distances was motivated by its superior behavior for systems with
long-ranged forces. The ZH closure can be rewritten as the following
expression for $T$ in the out region:

\begin{equation}
T^{ZH}(r)=\frac{\exp \{F_{\alpha }(r)[t(r)-\beta u_{1}(r)]\}-1-F_{\alpha
}(r)t(r)}{F_{\alpha }(r)}\;,  \label{TZH}
\end{equation}
where $F_{\alpha }(r)$ is an $r$-dependent interpolation function, 
\begin{equation}
F_{\alpha }(r)=1-\exp (-r/r_{\alpha }),  \label{FZH}
\end{equation}
and $r_{\alpha }$ a fitting parameter chosen to achieve thermodynamic
consistency. For $F_{\alpha }\rightarrow 0$ ( i.e. for $r/r_{\alpha
}\rightarrow 0)$ Eq.~(\ref{TZH}) reduces to the SMSA closure $-\beta u_{1}$
and for $F_{\alpha }\rightarrow 1$ ( i.e. for $r/r_{\alpha }\rightarrow
\infty )$ Eq.~(\ref{TZH}) reduces to the HNC closure, Eq.~(\ref{Thnc}),
though Eq.~(\ref{FZH}) implies a rather slow transition between these limits
for physically relevant values of $r_{\alpha }$.

For the systems we consider here with short-ranged interactions, the
important feature of Eq.~(\ref{TZH}) is not the behavior of the HNC equation
at large distances but the fact that at low densities $T^{HNC}$ reduces to
the exact result, Eq.~(\ref{Tlow}). ZH found numerically for the LJ fluid
that $r_{\alpha }$ decreased as the density tended to zero, so the HNC
closure is effectively used at all relevant $r$ at very low density. At
higher density $r_{\alpha }$ increases, thus mixing in more and more of the
SMSA expression. For example, near the triple point (at a reduced density of
0.85 and a reduced temperature of 0.786) ZH found that $r_{\alpha
}=6.25\sigma$\cite{hmsa}. The ZH interpolation scheme provides a mechanism
by which one can go between these limits as the density changes while
maintaining enough flexibility in the shape of $T$ outside the core that
thermodynamic consistency for the pressure can be achieved.

\subsection{Tail Interpolation Method}

\label{TIM}Both the ZH equation and the RHNC equation discussed above in
Sec.\ref{RHNC} give accurate results at both high and low densities by
considering some rather complicated density dependent expressions for the
out part of $T,$ which in particular involve $t(r).$ See Eqs.~(\ref{TZH})
and (\ref{Trhnc}). The variational method discussed in the Appendix can be
used to solve the OZ equation when the out part of $T$ is a known function
of $r$, as is the case for the SMSA approximation. Because of the appearance
of the initially unknown function $t(r)$ in the ZH and RHNC expressions for $%
T(r),$ we cannot use this method alone to solve these equations. However, by
making an initial guess for the out part of $T,$ we can iterate until the
value of the out part of $T$ does not change. This method combines the
standard Picard iteration scheme for the hopefully slowly varying out part
of $T$ with the efficient variational method for the core parts of
functions. While we have found that this method generally works quite well,
it still requires much more computer time than does the direct variational
solution of the OZ equation with a known out part of $T.$ Moreover because
of the complicated nonlinear nature of the self-consistency condition and
the direct interplay between possible oscillations in $t$ and $T$ in the out
region it is not clear that self-consistent solutions can always be found
for physically relevant states. Indeed the RHNC equation fails in a quite
peculiar way \cite{belloni1} close to the vapor-liquid coexistence region.

We now introduce a new method, which we call the {\em tail interpolation
method} (TIM),\ that implements the idea of a density dependent
interpolation involving $f_{1}$ and $-\beta u_{1}$ very directly, while
using a very simple ($t$ independent) expression for the out part of $T.$ We
assume the out part of $\Delta T$ can be written as: 
\begin{equation}
\Delta T^{TIM}(r)=\alpha f_{1}(r)+(1-\alpha )[-\beta u_{1}(r)],  \label{TTIM}
\end{equation}
where $\alpha $ is a (temperature and density dependent) parameter that is
chosen so that consistent results for two different routes to one particular
thermodynamic quantity are obtained. (To obtain the full $T$ the out part of 
$T_{0}$ should be added to Eq.~(\ref{TTIM}); often the SMSA-PY approximation 
$T_{0}=0$ gives sufficient accuracy.)\ Note that the presumably exact
asymptotic form for the tail of $c(r)$ given in Eq.~(\ref{climit}) is
maintained for {\em any} choice of $\alpha $, and $\alpha $ is not required
to lie between zero and one. In general, varying $\alpha $ allows us to
change the shape of the tail of $c$ at intermediate distances while
maintaining the proper asymptotic form, and we use this freedom to achieve
partial thermodynamic self-consistency. At low density $\alpha =1$ and,
given the relative accuracy of the SMSA, we expect that at high densities
smaller values of $\alpha $ will be found.

In this paper we impose consistency between the virial and compressibility
routes to the {\em isothermal compressibility}. Belloni \cite{belloni} has
shown that this can be implemented very efficiently by differentiating the
OZ equation, and our variational method can be easily extended to this case.
We have not yet examined in any detail the merits of this choice over other
possible consistency conditions. Indeed the SMSA usually gives rather poor
results both for the virial pressure and for the compressibility \cite
{MDHCYF}, and the energy route is typically used to give more accurate
thermodynamic results \cite{scoza2}. It is easy to derive a variational
method to impose thermodynamic consistency from the energy route and we
suspect this will give even better results. However, in this initial study
we have imposed consistency on the isothermal compressibility to see whether
self-consistency using the very simple expression for $T(r)$ given in Eq.~(%
\ref{TTIM}) can improve on the rather poor performance of the SMSA for this
quantity. The preliminary data we report in the next section illustrates the
basic concept and suggests that further work is indeed merited.

\section{Numerical Results}

We test our approach on two well-studied systems: the hard core Yukawa fluid
(HCYF) and the LJ fluid. The HCYF has been the focus of recent theoretical
work \cite{scoza2} and represents a system where errors from the treatment
of soft cores do not arise, while the LJ fluid is a typical soft core system.

\subsection{HCYF}

\begin{figure}
\epsfxsize=7in
\epsfbox{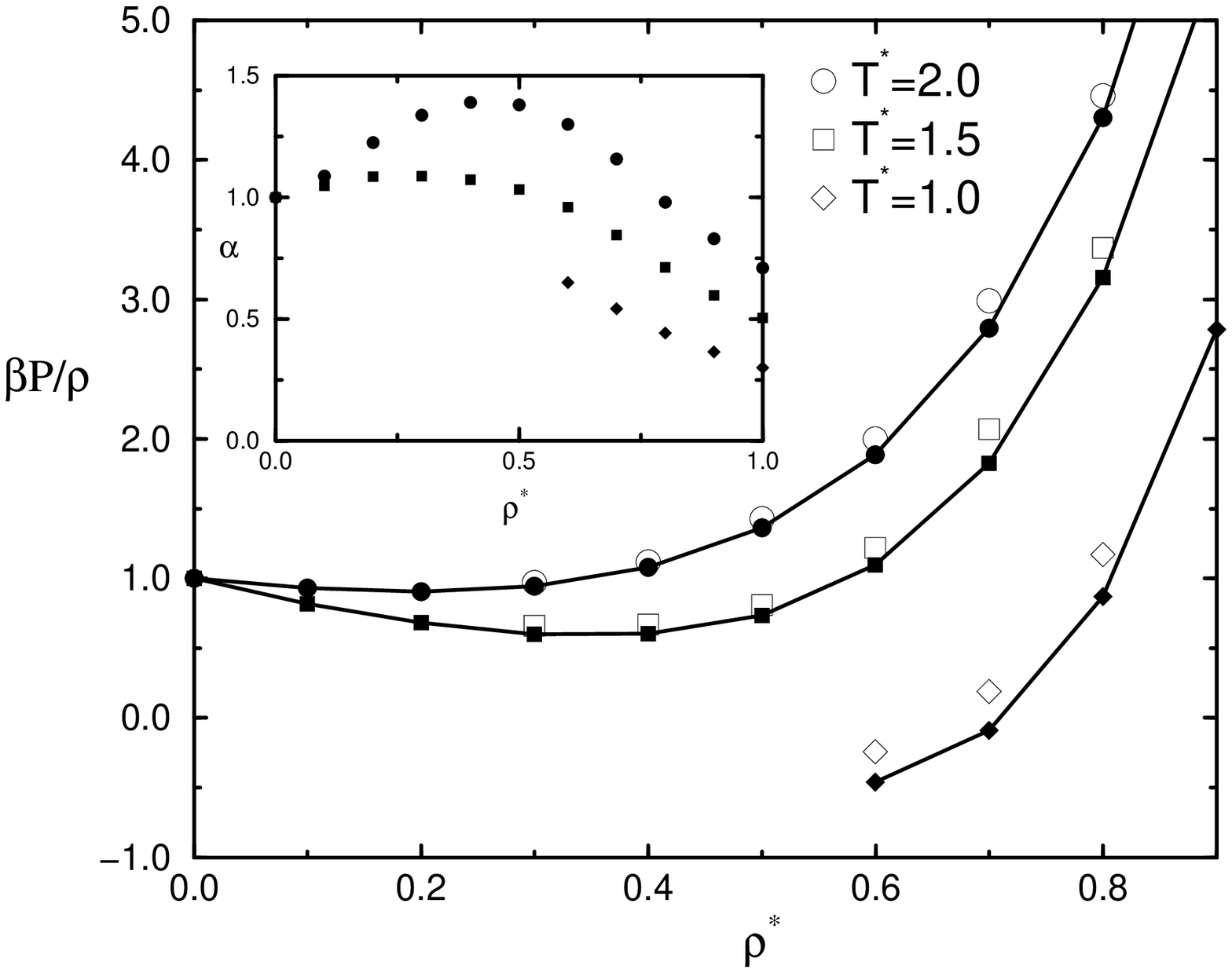}
\caption{Dependence of the compressibility factor $\beta P/\rho$
on density $\rho^*$ and temperature $T^*$ for the Yukawa fluid. Open
symbols represent the results of MD simulations \protect\cite{MDHCYF} and filled
symbols are the predictions of the TIM approximation. Lines are guides to the
eye. In the inset: dependence of the self-consistency parameter
$\alpha$ in the TIM
approach on density $\rho^*$ and temperature $T^*$.} 
\label{fig1}
\end{figure}

The interaction potential in the HCYF is given by: 
\begin{equation}
w^{HCYF}(r)=u_{d}(r)+\epsilon \frac{e^{-z(r-d)}}{r/d},  \label{wHCYF}
\end{equation}
where $d$ is the hard sphere diameter. We choose $z=1.8/d$, which corresponds
to a well-studied system \cite{MCHCYF,MDHCYF}. We have solved the TIM equations
using the variational method described below in the Appendix. For greater
accuracy in treating the hard sphere correlations at high density, we have
included a GMSA like expression for $T_{d}$ in the out region, as described
in the Appendix. Only preliminary results are reported here. In
Fig.~\ref{fig1} we 
give values for the compressibility factor $\beta P/\rho $ compared to the
results of a MD simulation study \cite{MDHCYF}. We emphasize that the
compressibility factor has been calculated directly from the virial formula
for pressure and not obtained through thermodynamic relations from the
energy route, as is usually done in ORPA and MSA approaches for greater
accuracy. In the inset to Fig.~\ref{fig1} we present the dependence of the TIM
self-consistency parameter $\alpha $ on temperature and density. Isotherms $%
T^{*}=2.0$ and $T^{*}=1.5$ are supercritical, and $T^{*}=1.0$ is subcritical 
\footnote{We use reduced units: $\rho ^{*}\equiv \rho d^{3}$, 
$T^{*}\equiv k_BT/\epsilon $, etc.}.
At low densities $\alpha $ approaches the exact limit $\alpha
=1$, while at higher densities $\alpha $ becomes smaller though differing
from zero (the MSA limit). The behavior at intermediate densities where $%
\alpha $ reaches a maximum is interesting and was not anticipated by us. The
behavior of the TIM very near the critical point and spinodal lines has not
been examined.

\begin{figure}
\epsfxsize=7in
\epsfbox{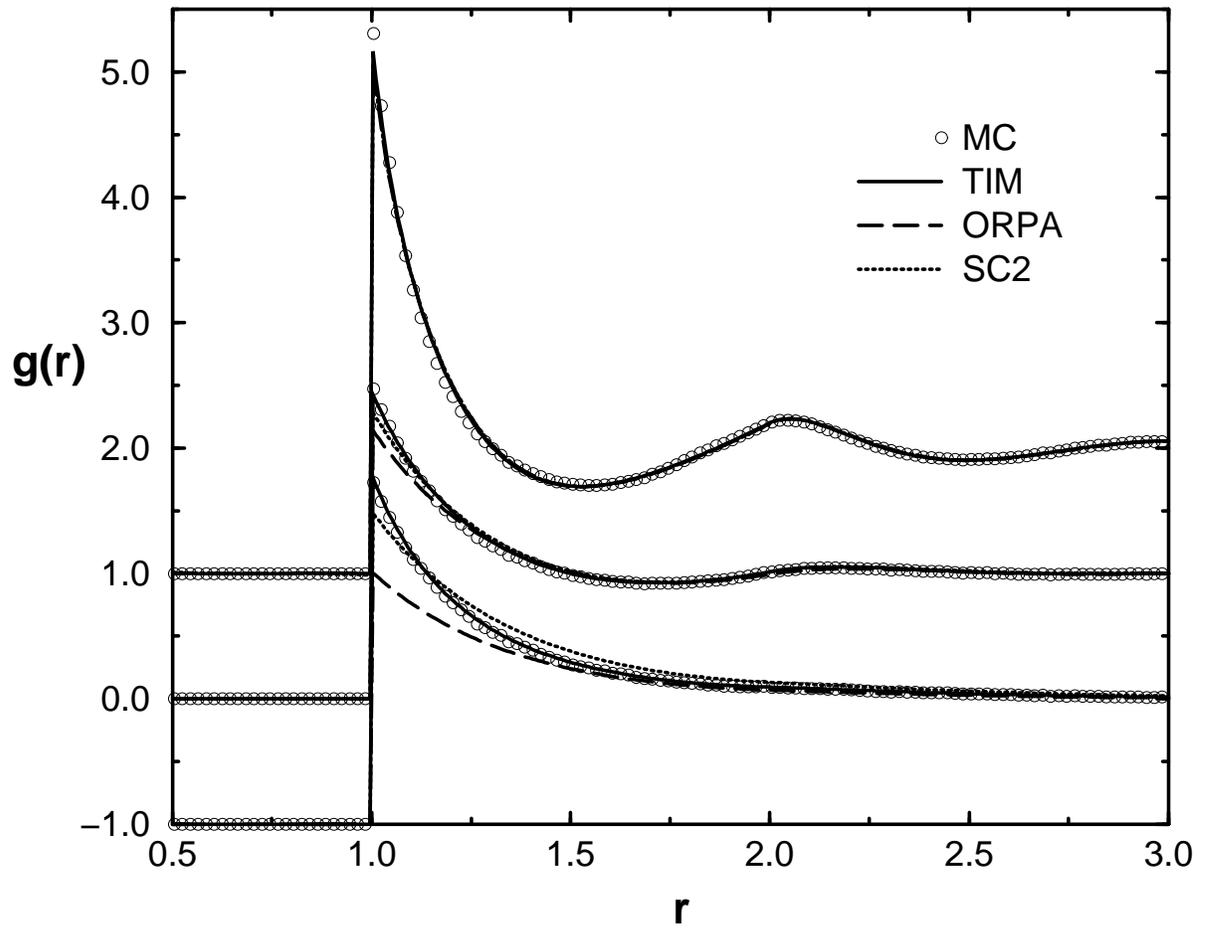}
\caption{Density correlation functions of the hard core Yukawa fluid. From top
to bottom: ($\rho^*=0.8, T^*=0.9$), ($\rho^*=0.4,T^*=1.25$),
($\rho^*=0.05, T^*=1.0$). MC simulations performed in this work. For
the sake of clarity, curves have been shifted in the 
vertical direction.}
\label{fig2}
\end{figure}

To test the accuracy of the correlation functions predicted by the TIM, we
compare them to the results of new MC simulations we have carried out 
\footnote{This is a standard NVT-ensemble MC simulation, with the
number of particles ranging from 128 to 432, depending on density. The pair
potential has been cut and shifted at $r_{cut}/d=3$.}. 
In Fig.~\ref{fig2} we show $h(r)$ given by the TIM, the ORPA, and an even
simpler self-consistent approach (SC2) very similar to that used by Stell
and coworkers \cite{scoza2}, where $T^{SC2}(r)=\alpha [-\beta u_{1}(r)]$,
with $\alpha $ is chosen to satisfy self-consistency of the virial and
compressibility routes to the compressibility. Since the SC2 tail does not
have enough flexibility to reduce to $f_{1}$ at low density, we expect that
its correlation functions may be less accurate there. The results show the
relatively inaccuracy of the ORPA at intermediate and low densities, with
best results seen at high density. The SC2 approach, while giving accurate
self-consistent thermodynamics, yields less accurate correlation functions
at low densities, as expected.

\subsection{LJ fluid}

\begin{figure}
\epsfxsize=7in
\epsfbox{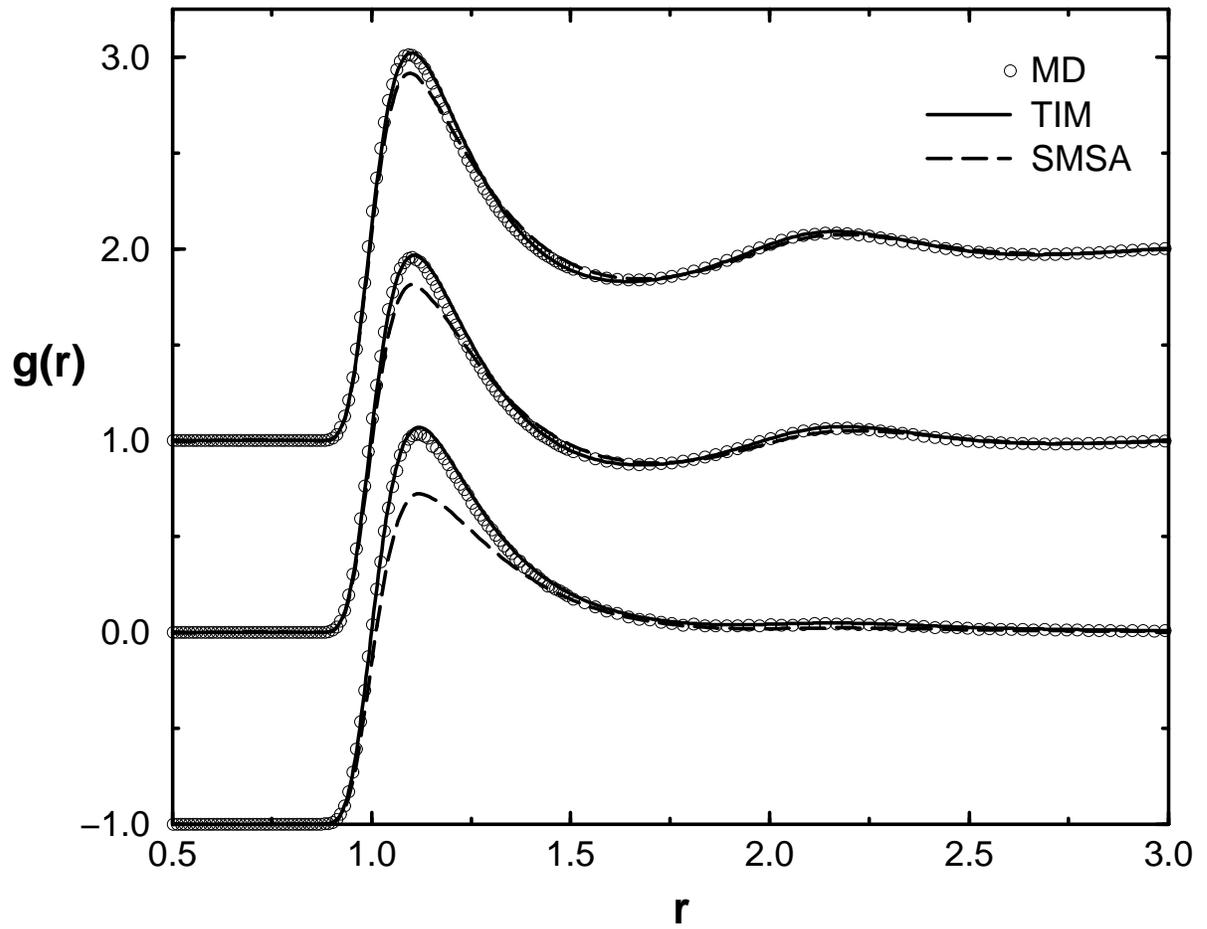}
\caption{Density correlation functions of the Lennard-Jones fluid. From top
to bottom: ($\rho^*=0.54, T^*=1.35$), ($\rho^*=0.45,T^*=1.35$),
($\rho^*=0.1, T^*=1.35$). MD simulations are taken from \protect\cite{MDWVK}. For
the sake of clarity, curves have been shifted in the
vertical direction.}
\label{fig3}
\end{figure}

We have also solved the TIM equations for the LJ fluid for a few states,
using the WCA separation of the pair potential \cite{wca}. For the
relatively low density states we study here, the SMSA-PY approximation for
the reference system $T_{0}=0$ gives sufficient accuracy. In Fig.~\ref{fig3} we
compare predictions of TIM and SMSA approximations to MD simulations results
\cite{MDWVK}. The states shown correspond to low and moderate densities at
about the critical temperature. Again the TIM approach gives notable
improvement over the SMSA theory, especially at low densities.

\section{Final Remarks}

Many issues in the theory of integral equations for fluid structure can be
profitably analyzed and interpreted in terms of predictions for the out part
of the tail function $T(r),$ i.e., the tail of the direct correlation
function. In this paper we have only considered a single component uniform
fluid with short ranged interactions. Here the simplest possible MSA linear
response approximation relating the tail of $T(r)$ to the perturbation
potential $u_{1}(r)$ immediately yields the SMSA
theory. For systems with harshly repulsive forces only, the SMSA reduces to
the successful PY theory. For systems with a weak potential tail $u_{1}(r),$
the SMSA gives rather accurate results at high density but fails at low
density. The behavior at low densities can be greatly improved by
introducing a density dependence into the tail of $T(r)$ such that it
reduces to the exact low density result of Eq.~(\ref{Tlow}), as is
effectively done in the RHNC and HMSA equations. We introduced here a new
self-consistent (TIM) method that incorporates this idea in a much simpler
form, and the preliminary results for the LJ fluid and the HCYF appear
promising.

The success of all these methods at high density arises from the fact that
for the systems considered attractive forces have only a relatively small
effect on the liquid structure, so that $h(r)\approx h_{0}(r)$ is a fairly
good approximation. Put another way, the density fluctuations can be well
described by a simple Gaussian theory \cite{gaus}. When this is not 
true, as is the case
for nonuniform liquids \cite{wkv}, particularly in cases of wetting and drying
phenomena, the natural (singlet) generalizations of all these integral
equation methods fail\cite{sullstell}. We simply do not know a good 
enough guess for the
tail of $T(r)$ in cases where attractive forces induce such significant
structural changes. New approaches based on a self-consistent mean field
treatment of the attractive interactions appear more promising here \cite{wkv}.

A more severe test of these ideas and of the utility of integral equations
in general for uniform fluids is in applications to systems with long-ranged
Coulomb interactions. Here one must deal with fluid mixtures with strong and
long-ranged attractive and repulsive interactions, and the correlation
functions differ in significant ways from those of any reference system with
short-ranged forces. Nevertheless, characteristic properties of systems with
long-ranged forces, such as the Stillinger-Lovett moment conditions are very
naturally expressed in terms of the tail of the direct correlation
function \cite{hansenmac}.
The RHNC and HMSA approximations have often proved useful here, and even the
simple SMSA captures Debye screening, perhaps the most fundamental feature
of the long-ranged force problem. We hope that some the ideas presented here
for the $T$ function can be extended to long-ranged force systems to provide
a more intuitive understanding of the strengths and weaknesses of existing
integral equation approaches, and aid in the development of new and simpler
approximations.

\section{Acknowledgments}

It is a pleasure to dedicate this paper to George Stell on the happy
occasion of his 65th birthday. This work was supported by NSF Grant No.
CHE9528915.

\appendix
\section{Variational Method}

\subsection{Fixed Tail Function}

We can look on the OZ equation (\ref{OZ}) as indirectly relating the
continuous functions $T(r)$ and $t(r).$ Thus, given $T(r)$ we can in
principle solve for $t(r)$ and then determine $h(r)$ and $c(r)$ from Eqs.~(%
\ref{cSinout}) and (\ref{gSD}). We first consider the simplest case,
exemplified by the SMSA, where the out part of $T(r)$ is a fixed prescribed
function, independent of other correlation functions and/or the density.
Then we show how to generalize this approach for an arbitrary choice of $%
T(r) $, which can be coupled to other correlation functions such as $t(r).$
In the latter application our approach represents a new way to solve
standard integral equations, and we believe it offers some notable
advantages over conventional methods.

It is easy to rewrite the OZ equation (\ref{OZ}) in terms of $t$ and $c.$
Taking Fourier transforms we have 
\begin{equation}
\hat{t}(k)=\frac{\rho \hat{c}^{2}(k)}{1-\rho \hat{c}(k)},  \label{that}
\end{equation}
where $\hat{c}(k)$ denotes the Fourier transform of $c(r).$ As noted above,
only the ``out projection'' $e_{0}(r)T(r)$ is actually relevant for $h$ and $%
c$. Given this, Eq.~(\ref{cSinout}) shows that we need to fix only the
``core projection'' $f_{0}(r)t(r)$ to determine $c(r)$ for all $r$. In
principle, $t(r)$ can then be determined everywhere from the modified OZ
equation (\ref{that}). A proper self-consistent choice for $t(r)$ inside the
core must yield the same functional form when it is computed indirectly
using the OZ equation (\ref{that}). This requirement can be formulated very
efficiently in terms of a variational procedure.

In the following analysis $e_{0}(r)T(r)$ is held constant and variations in
all functions are generated solely by variations in $t(r)$ restricted to the
core region $r<\bar{\sigma}$. According to Eq.~(\ref{cSinout}), variations
of $c(r)$ and $t(r)$ then are linearly related: 
\begin{equation}
\delta c(r)=f_{0}(r)\delta t(r).  \label{deltac}
\end{equation}
To arrive at the proper variational functional, we first formally integrate
the r.h.s.~of Eq.~(\ref{that}) with respect to $\hat{c}$, thus arriving at a
functional 
\begin{equation}
\Phi _{OZ}=-\frac{1}{(2\pi )^{3}}\int \{\rho ^{2}\hat{c}^{2}(k)/2+\rho \hat{c%
}(k)+\ln [1-\rho \hat{c}(k)]\}d{\bf k\;,}  \label{Foz}
\end{equation}
whose general variation with respect to $\hat{c}$ can be simplified using
the modified OZ equation (\ref{that}): 
\begin{eqnarray}
\delta \Phi _{OZ} &=&-\frac{1}{(2\pi )^{3}}\int \left[ \rho ^{2}\hat{c}%
(k)+\rho -\frac{\rho }{1-\rho \hat{c}(k)}\right] \delta \hat{c}(k)d{\bf k} 
\nonumber \\
&=&\frac{\rho ^{2}}{(2\pi )^{3}}\int \hat{t}(k)\delta \hat{c}(k)d{\bf k\;.}
\label{delFcoz}
\end{eqnarray}
Using Parseval's formula, and considering the special variation of $c$ given
by Eq.~(\ref{deltac}), we have 
\begin{equation}
\delta \Phi _{OZ}=\rho ^{2}\int f_{0}(r)t(r)\delta t(r)d{\bf r\;,}
\label{delFtoz}
\end{equation}
which expresses the result in terms of the imposed variation of $t$ inside
the core. In Eq.~(\ref{delFtoz}), $t(r)$ satisfies the OZ equation (\ref
{that}). We now consider a second functional of $t$ : 
\begin{equation}
\Phi _{dir}=-\frac{\rho ^{2}}{2}\int f_{0}(r)t^{2}(r)d{\bf r}  \label{Fdir}
\end{equation}
whose variation {\em directly} gives the negative of the r.h.s.~of Eq.~(\ref
{delFtoz}). Thus by construction, the functional

\begin{equation}
\Phi \equiv \Phi _{OZ}+\Phi _{dir}  \label{Ftotal}
\end{equation}
obtained by adding Eqs.~(\ref{Foz}) and (\ref{Fdir}) is stationary (and
reaches its minimum) when the proper self-consistent value for $t(r)$ inside
the core is used.

To implement this variational method, we expand the core part of $t(r)$ for $%
r<\bar{\sigma}$ in terms of Legendre polynomials, orthogonal on $[0,\bar{%
\sigma}]$:

\begin{equation}
t(r)=\sum_{i=1}^{n}a_{n}P_{n}(r),\;\;r<\bar{\sigma}.  \label{tpoly}
\end{equation}
We choose values of the coefficients $a_{n}$ to minimize the functional $%
\Phi $ in Eq.~(\ref{Ftotal}). We have used Powell's quadratically convergent
method to implement the minimization procedure \cite{NumRec}. If needed, one
could improve this step of the calculation by using conjugate gradient
methods. In practice, our implementation is very efficient, and $t$ is
smooth enough that it is generally sufficient to use $n\approx 5$ to get
highly accurate results. For example, for hard core systems $-(1+t)=c$
inside the core, and the exact solution of PY equation for hard spheres
gives a $c$ that is a cubic polynomial \cite{hansenmac}. Fast Fourier
Transform methods \cite{NumRec} are used in evaluating Eq.~(\ref{Foz}).

\begin{figure}
\epsfxsize=7in
\epsfbox{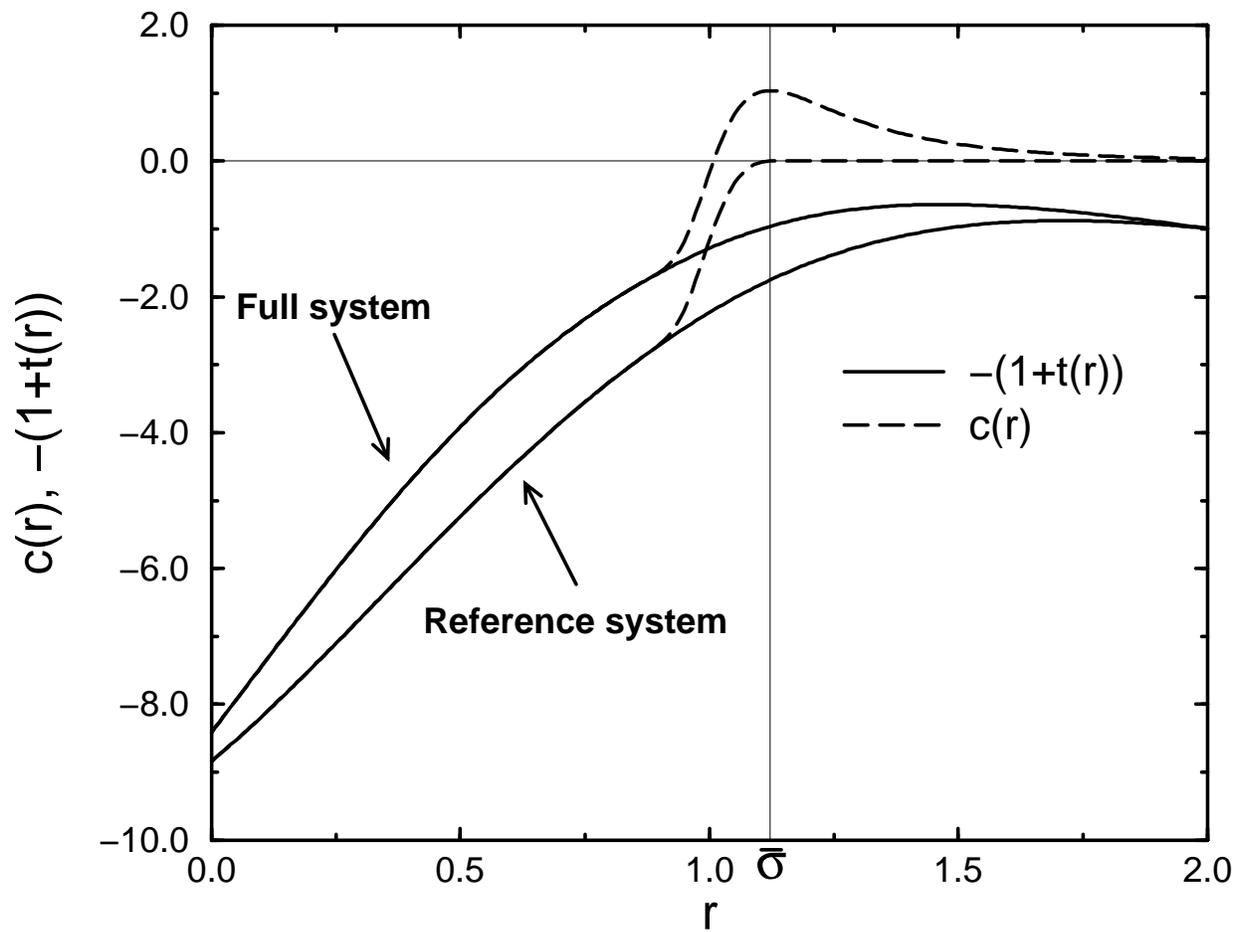}
\caption{Comparison of the shape of $c(r)$ and $t(r)$ for the
reference and full Lennard-Jones fluid ($\rho^*=0.54, T^*=1.35$). This
shows the advantage of using $t(r)$ as a variational function instead of
$c(r)$ as used in \protect\cite{narten}.}
\label{fig4}
\end{figure}

One important general feature of our method is that we expand the smooth
function $t$ inside the core rather than $c.$ As discussed above, for hard
core systems these two procedures are equivalent, and our method then
reduces exactly to the variational method Andersen and Chandler \cite{ac72}
used to solve the hard core MSA and ORPA equations. However, for soft core
systems, while $c(r)$ is simply related to $t(r)$ well inside the core, it
changes rapidly in the narrow transition region, close to $r\approx \bar{%
\sigma}$. Higher order polynomials are required to describe this rapid
localized variation of $c$ accurately. This problem becomes more and more
severe with increasing steepness of the reference system potential. However $%
t$ changes smoothly and slowly even in the transition region. This is
illustrated in Fig.~\ref{fig4}, where we show $t(r)$ and $c(r)$ both for the full
LJ system, calculated using the self-consistent TIM method discussed in
Sec.~(\ref{TIM}), and for the LJ reference system using the SMSA (PY)
approximation. The same qualitative features are seen both at lower and
higher densities and using other accurate closures. Previous variational
methods proposed for soft core systems \cite{narten,olivares} have either
expanded $c(r)$ inside the core or $h(r)$ itself.

\subsection{Arbitrary Tail Function}

To solve integral equations for an arbitrary prescription for $T(r)$
containing initially unknown functions like $t(r)$ (see, e.g., Eqs.~(\ref{py}%
), (\ref{Thnc}), and (\ref{Trhnc}) for the PY, HNC and RHNC equations) we
can combine the variational technique with an iterative method. First, we
make an initial guess $T^{(0)}(r)$ for the out part of $T$, and solve the
variational problem as above for this fixed choice. This will yield new
values for $t$ and other correlation functions. Then, the next approximation 
$T^{(1)}(r)$ can be determined for a given closure and the obtained
correlation functions. The new approximation is used in the next variational
step and this iteration procedure is repeated until convergence of the tail
of $T(r)$ is obtained. Again one could replace the iterative steps by more
sophisticated methods \cite{gillan}. However, since the out part of $T$ has
a relatively simple structure, we have found the simple iterative method
works quite well in all cases we have tested.

\subsection{Thermodynamic self-consistency}

By introducing a dependence of $T(r)$ on one free parameter we can ensure
(partial) self-consistency of thermodynamic properties. Here we extend our
variational method to impose consistency between the virial and
compressibility routes to the isothermal compressibility: 
\begin{equation}
\chi _{T}^{V}(\rho ,\beta )=\chi _{T}^{C}(\rho ,\beta ).  \label{consistency}
\end{equation}
Here 
\begin{equation}
\chi _{T}^{V}(\rho ,\beta )=\left( \frac{\partial \beta P^{V}}{\partial \rho 
}\right) _{\beta }=1-\frac{\partial }{\partial \rho }\left( \frac{\beta \rho
^{2}}{6}\int r\frac{dw(r)}{dr}g(r)d{\bf r}\right) ,  \label{chiV}
\end{equation}
\begin{equation}
\chi _{T}^{C}(\rho ,\beta )=\left( \frac{1}{1-\rho \hat{c}(k)}\right)
_{k=0}\;.  \label{chiC}
\end{equation}
To evaluate $\chi _{T}^{V}(\rho ,\beta )$ we need an efficient way to
calculate $\partial g(r)/\partial \rho .$ Just as we did for $g(r)$ we can
use a variational method.

To simplify notation, let us denote the density derivative of a function $%
f(r;\rho )$ as $f_{\rho }^{\prime }(r)\equiv \partial f(r;\rho )/\partial
\rho .$ The OZ equation (\ref{that}) and relation (\ref{cSinout}) can be
directly differentiated: 
\begin{equation}
\hat{t}_{\rho }^{\prime }(k)=\frac{\hat{c}^{2}(k)+\rho \hat{c}(k)[2-\rho 
\hat{c}(k)]\hat{c}_{\rho }^{\prime }(k)}{[1-\rho \hat{c}(k)]^{2}}\;,
\label{DOZE}
\end{equation}
\begin{equation}
c_{\rho }^{\prime }(r)=f_{0}(r)t_{\rho }^{\prime }(r)+e_{0}(r)T_{\rho
}^{\prime }(r),  \label{DcSinout}
\end{equation}
\begin{equation}
g_{\rho }^{\prime }(r)=t_{\rho }^{\prime }(r)-c_{\rho }^{\prime }(r).
\label{Dg}
\end{equation}

To solve (\ref{DOZE}) and (\ref{DcSinout}) one needs to know the function $%
T_{\rho }^{\prime }(r)$. In the simplest approach, this derivative is
neglected, because its density dependence is usually very weak. This has
been referred to as {\em local consistency} \cite{belloni}. Alternatively,
one can first calculate $T(r)$ with $T_{\rho }^{\prime }(r)$ set to zero and
then compute its density derivative by a finite difference method by
evaluating $T(r)$ at slightly different densities. The results can be
plugged back into (\ref{DcSinout}) and the equations iterated until
convergence to {\em globally consistent} results, like those in Ref.\cite
{scoza2}, are found.

In calculations of correlation functions of the HCYF we have introduced a
small correction to the out part of $T(r)$, which corresponds to a proper
description of the pure hard-sphere system (the limiting case of HCYF as $%
\beta \to 0$). This correction has the usual GMSA--like Yukawa form \cite
{wais} 
\begin{equation}
T_{d}(r)=K_{d}(\rho )\exp [-z_{d}(\rho )(r-d)]/r,
\end{equation}
with $K_{d}(\rho )$ and $z_{d}(\rho )$ chosen to satisfy consistency of the
virial and compressibility pressure and the Carnahan-Starling equation of
state. In practice, we used results of Tang and Lu \cite{tanglu}, who
derived very accurate (approximate) explicit analytic expressions for $%
K_{d}(\rho )$ and $z_{d}(\rho )$. Using these, one can explicitly evaluate
the density derivative of $T_{d}(r),$ thus ensuring global self-consistency
for the hard core reference system of the HCYF.

%
%

\newpage
%
%

\end{document}